\documentclass[twocolumn,showpacs,preprintnumbers,amsmath,amssymb,prl,superscriptaddress,floatfix]{revtex4}

\usepackage{graphicx}
\usepackage{dcolumn}
\usepackage{bm}
\usepackage{amsfonts}
\usepackage{amsmath}
\usepackage{amssymb}
\newcommand{\mr}{\mathrm}

\begin{document}
\title{Observation of inverse diproton photodisintegration at intermediate energies}
\author{V.~Komarov}
\affiliation{Laboratory of Nuclear Problems, Joint Institute for Nuclear
Research, 141980 Dubna, Russia}
\author{T.~Azaryan}
\affiliation{Laboratory of Nuclear Problems, Joint Institute for
Nuclear Research, 141980 Dubna, Russia}
\author{D.~Chiladze}\affiliation{Institut f\"ur Kernphysik, Forschungszentrum J\"ulich,
52425 J\"ulich, Germany}%
\affiliation{High Energy Physics Institute, Tbilisi State University,
0186 Tbilisi, Georgia}
\author{S.~Dymov}\email{s.dymov@fz-juelich.de}%
\affiliation{Laboratory of Nuclear Problems, Joint Institute for
Nuclear
Research, 141980 Dubna, Russia}%
\affiliation{Physikalisches Institut II, Universit{\"a}t
Erlangen-N{\"u}rnberg, 91058 Erlangen, Germany }
\author{M.~Hartmann}
\affiliation{Institut f\"ur Kernphysik, Forschungszentrum J\"ulich,
52425 J\"ulich, Germany}
\author{A.~Kacharava}\affiliation{Institut f\"ur Kernphysik, Forschungszentrum
J\"ulich, 52425 J\"ulich, Germany}
\author{I.~Keshelashvili}\affiliation{Institut f\"ur Kernphysik, Forschungszentrum J\"ulich,
52425 J\"ulich, Germany}%
\affiliation{High Energy Physics Institute, Tbilisi State University,
0186 Tbilisi, Georgia}
\author{A.~Khoukaz}
\affiliation{Institut f\"ur Kernphysik, Universit\"at M\"unster, 48149 M\"unster, Germany}%
\author{A.~Kulikov}%
\affiliation{Laboratory of Nuclear Problems, Joint Institute for
Nuclear Research, 141980 Dubna, Russia}
\author{V.~Kurbatov}
\affiliation{Laboratory of Nuclear Problems, Joint Institute for
Nuclear Research, 141980 Dubna, Russia}
\author{G.~Macharashvili}
\affiliation{Laboratory of Nuclear Problems, Joint Institute for
Nuclear Research, 141980 Dubna, Russia}%
\affiliation{High Energy Physics Institute, Tbilisi State University,
0186 Tbilisi, Georgia}
\author{S.~Merzliakov}
\affiliation{Laboratory of Nuclear Problems, Joint Institute for
Nuclear Research, 141980 Dubna, Russia}%
\affiliation{Institut f\"ur Kernphysik, Forschungszentrum J\"ulich,
52425 J\"ulich, Germany}
\author{S.~Mikirtychiants}
\affiliation{Institut f\"ur Kernphysik, Forschungszentrum J\"ulich,
52425 J\"ulich, Germany}%
\affiliation{High Energy Physics Department, Petersburg Nuclear
Physics Institute, 188350 Gatchina, Russia}
\author{M.~Papenbrock}
\affiliation{Institut f\"ur Kernphysik, Universit\"at M\"unster,
48149 M\"unster, Germany}%
\author{M.~Nekipelov}
\affiliation{Institut f\"ur Kernphysik, Forschungszentrum J\"ulich,
52425 J\"ulich, Germany}%
\author{F.~Rathmann}
\affiliation{Institut f\"ur Kernphysik, Forschungszentrum J\"ulich,
52425 J\"ulich, Germany}
\author{V.~Serdyuk}%
\affiliation{Laboratory of Nuclear Problems, Joint Institute for
Nuclear Research, 141980 Dubna, Russia}%
\affiliation{Institut f\"ur Kernphysik, Forschungszentrum J\"ulich,
52425 J\"ulich, Germany}
\author{H.~Str\"oher}\affiliation{Institut f\"ur Kernphysik, Forschungszentrum
J\"ulich, 52425 J\"ulich, Germany}
\author{D.~Tsirkov}
\affiliation{Laboratory of Nuclear Problems, Joint Institute for
Nuclear Research, 141980 Dubna, Russia}%
\author{Yu.~Uzikov}\affiliation{Laboratory of Nuclear Problems, Joint Institute for Nuclear
Research, 141980 Dubna, Russia}
\author{C.~Wilkin}\affiliation{Physics and Astronomy Department, UCL,
London, WC1E 6BT, UK}
\date{\today}

\begin{abstract}
The reaction $pp\rightarrow\{pp\}_{\!s}\gamma$, where $\{pp\}_{s}$ is
a proton pair with an excitation energy $E_{pp}<3$~MeV, has been
observed with the ANKE spectrometer at COSY-J\"{u}lich for proton
beam energies of $T_p=0.353$, 0.500, and 0.550~GeV. This is
equivalent to photodisintegration of a free $^{1\!}S_0$ diproton for
photon energies $E_{\gamma}\approx T_p/2$. The differential cross
sections measured for c.m.\ angles $0^{\circ}<\theta_{pp}<
20^{\circ}$ exhibit a steep increase with angle that is compatible
with $E1$ and $E2$ multipole contributions. The ratio of the measured
cross sections to those of $np\rightarrow d\gamma$ is on the
$10^{-3}-10^{-2}$ level. The increase of the
$pp\to\{pp\}_{\!s}\gamma$ cross section with $T_p$ might reflect the
influence of the $\Delta(1232)$ excitation.
\end{abstract}


\maketitle

Photoabsorption on two-nucleon systems in nuclei at several
hundred MeV allows one to probe fundamental properties of nuclei
at short distances. The photodisintegration of the simplest
nucleus, the deuteron, through the $\gamma d\rightarrow pn$
reaction is widely used as a testing ground for different
theoretical ideas of the nucleon-nucleon interaction, such as
meson-exchange models and isobar currents~\cite{Arenhoevel} or,
more recently, quark-gluon degrees of freedom~\cite{Gilman}.
However, much less is known, both experimentally and
theoretically, on the other simplest process
\begin{equation}
\gamma+\{pp\}_{\!s}\rightarrow p+p\,, \label{gammapps}
\end{equation}
where $\{pp\}_{\!s}$ is a proton pair in the $^{1\!}S_0$ state. The
photodisintegration of the spin-singlet \textit{pp}-pair differs from
that of the spin-triplet ($^{3\!}S_1-\,^{3\!}D_1$) $pn$ pair, where
the $M1$ magnetic dipole transition dominates $\gamma d\rightarrow
pn$ at several hundred MeV through the excitation of the
$\Delta(1232)$ isobar~\cite{Crawford,Wilhelm}. In contrast, due to
selection rules, there is no direct contribution to reaction
(\ref{gammapps}) from $S$-wave $\Delta N$ intermediate
states~\cite{Laget,WilhelmPR} and $M$-odd multipoles are forbidden.
Furthermore, since the diproton has no electric dipole moment, only
the spin-flip contribution to the $E1$ operator
survives~\cite{WilhelmPR}. Features of the underlying dynamics, which
are not visible in the photodisintegration of the deuteron, may
therefore reveal themselves in reaction (\ref{gammapps}).

In the absence of a free bound diproton, reaction (\ref{gammapps})
has only been investigated for a $^{1\!}S_0$ diproton bound within a
nucleus, the lightest of these being
$^3\mr{He}$~\cite{Emura,Niccolai,Audit}. However, since the $M1$
absorption on quasi-deuteron pairs is so strong, the
$^3\mr{He}(\gamma,pp)n$ reaction has large backgrounds associated
with apparent three-nucleon absorption, combined with final state
interactions (FSI). The total cross section for photon absorption by
two protons in $^3\mr{He}$ for photon energies 0.2 -- 0.5~GeV was
found to be only a few percent of the total rate~\cite{Emura}. These
contaminations are absent in the inverse reaction with the production
of a free $^{1\!}S_0$ diproton
\begin{equation}
p+p\rightarrow \gamma+\{pp\}_{\!s}\,. \label{eq2}
\end{equation}
At excitation energies $E_{pp}$ of the final $pp$ pair less than a
few MeV ($E_{pp}<3$~MeV, for definiteness), the system is almost
exclusively in the $^{1\!}S_0$ state.

The known experiments on hard $pp$ bremsstrahlung at intermediate
energies were not designed for the study of the quasi-two-body
channel (\ref{eq2}). In the published
data~\cite{NefkensPRC,Yasuda,Michaelian,Przewoski,Zlomanchuk}, the
selection of low $E_{pp}$ events was either impossible instrumentally
or was not done if feasible. In the COSY--TOF experiment at a beam
energy of $T_p=0.293$~GeV, the $pp\gamma$ data did not exhibit any
sizeable FSI enhancement at low $E_{pp}$ and no estimate of the cross
section for channel (\ref{eq2}) was made~\cite{Bilger}. The aim of
the present work was to observe the reaction in the region of the
$\Delta$ and to measure its differential cross section. Here we
present results at $T_p=0.353$, 0.500, and 0.550~GeV.

The experiment was performed using the ANKE setup~\cite{Barsov}
installed at the internal proton beam of the synchrotron storage ring
COSY-J\"{u}lich. Positively charged secondaries produced in the
hydrogen cluster-jet target traversed the vertical magnetic field of
the spectrometer and entered the forward detector, consisting of
multiwire chambers followed by a hodoscope of vertically oriented
scintillators.

\begin{figure}[t]
\includegraphics[width=0.48\textwidth]{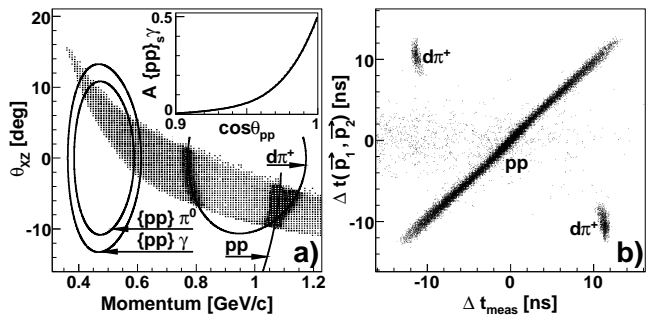}
\caption{Performance of the setup at 0.5~GeV. a) The polar angle
projection $\theta_{XZ} $ onto the median plane of the spectrometer
plotted versus the particle momentum. The experimental points show
the acceptance for detection of single particles. The curves depict
the kinematical loci for $p$ and $d$ from the indicated processes.
The symbol $\{pp\}$ denotes a \textit{pp} pair with the invariant
mass equal to $2m_{p}$. The acceptance $A{\{pp\}_{\!s}}\gamma$ for
reaction~(\ref{eq2}) is shown in the inset as a function of the c.m.\
polar angle $\theta_{pp}$ of the proton pair. b) Identification of
the proton pairs, as described in the text.} \label{fig1}
\end{figure}

The acceptance of the system is shown in Fig.~1a in terms of the
particle momentum and the projection $\theta _{XZ}$ of the emission
angle onto the horizontal plane. It is seen that the setup allows the
recording of protons from reaction~(\ref{eq2}) for
$|\theta_{XZ}|\lesssim 5^{\circ}$. The vertical acceptance is $\pm
3.5^{\circ}$. The acceptance for the proton pairs peaks at small
polar angles $\theta_{pp}$ (inset in Fig.~\ref{fig1}a).

Single protons from elastic \textit{pp} scattering, identified by
their momenta and ionization losses, were recorded for luminosity
purposes. Differential cross sections from Ref.~\cite{Arndt} were
used to establish the absolute normalization. The number of protons
stored in the ring was typically a few times $10^{10}$ and the target
density was $10^{14}$ protons/cm$^2$. The errors in the integrated
luminosities in Table~\ref{tab1} include both systematic and
normalization effects. More detailed descriptions of the setup and
data-processing procedure can be found
elsewhere~\cite{DymovPN,DymovPLB,Kurbatov}.

\begin{table}[htb]
\caption{Measurement characteristics: $L_\mr{int}$  is the integrated
luminosity; $(M_{x}^{\,2})^m$ and $\sigma(M_x^{\,2})$ in
$0.01~\rm{GeV}^2/c^4$ units are, respectively, the mean value and the
standard deviation of the missing-mass-squared distributions for $pp$
pairs with $E_{pp}<3$~MeV at the beam energy $T_{p}$; $N_{\gamma}$ is
the number of events in the $\gamma$-peak for $\theta_{pp}<20^\circ$;
$N_\mr{bg}/N_\gamma$ is the ratio of the background to signal in the
$\gamma$ peak; $E_x^{m}$ is the mean value of the missing c.m.\
energy for the events from the $\gamma$ peak; $E_\gamma$ is the
laboratory energy of the photon in the inverse
reaction~(\ref{gammapps}). }
\begin{ruledtabular}
\begin{tabular}{cccc}
$T_p$ [GeV]                & 0.353         & 0.500         & 0.550         \\ %
$E_\gamma$ [GeV]           & 0.176         & 0.249         & 0.274        \\ \hline %
$L_\mr{int}$ [nb$^{-1}$]   & 573$\pm$18    & 331$\pm$10    & 318$\pm$21    \\ \hline %
$(M_x^{\,2})^m_\gamma$     & 0.01$\pm$0.03 & 0.02$\pm$0.04 & 0.01$\pm$0.04 \\
$\sigma(M_x^2)_\gamma$     & 0.28$\pm$0.04 & 0.35$\pm$0.03 & 0.41$\pm$0.02 \\
$N_\gamma$                 & 180           & 335           & 525           \\
$N_\mr{bg}/N_\gamma$       & 0.23          & 0.05          & 0.11          \\
$E_x^m$ [GeV]              & 0.161         & 0.221         & 0.241         \\
\end{tabular}
\end{ruledtabular}
\label{tab1}
\end{table}

When proton pairs hit different counters, the difference $\Delta
t_\mr{meas}$ of the arrival times can be measured and compared with
the time-of-flight difference $\Delta t(\vec{p}_1,\vec{p}_2)$
calculated using the measured momenta, assuming that both particles
are protons. The $\Delta t_\mr{meas}-\Delta t(\vec{p}_1,\vec{p}_2)$
distribution has a FWHM of 0.6 -- 1.1~ns, so that genuine proton
coincidences can be identified unambiguously (Fig.~\ref{fig1}b). The
tracking system led to a precision in the determination of the
momentum $\sigma(p)/p\approx1\%$ and polar angle
$\sigma(\theta)\approx0.2^{\circ}$ for protons around 0.6~GeV/$c$ and
these gave a resolution $\sigma(E_{pp})=0.1-0.5$~MeV for
$E_{pp}<3$~MeV.

\begin{figure}[b]
\includegraphics[width=0.48\textwidth]{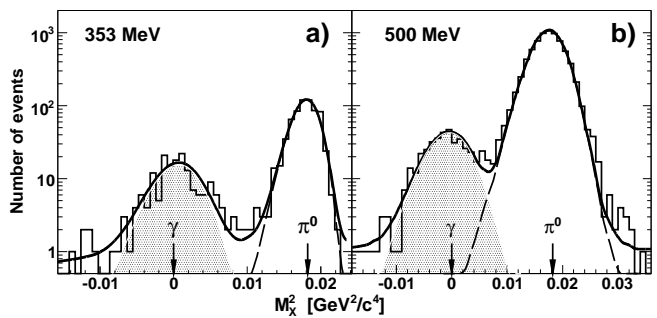}
\caption{Missing-mass-squared distributions for the $pp\to ppX$
reaction for events with $E_{pp} < 3$~MeV. The shaded area
corresponds to the predicted $\gamma$ peak, the dashed line to the
$\pi^0$, and the solid to the sum of these and a straight line
background.} \label{fig2}
\end{figure}

The determination of the four-momenta of the two final protons allows
a full kinematical reconstruction of $pp\rightarrow ppX$ events and
the derivation of the missing-mass spectra for the pairs with
$E_{pp}<3$~MeV. In our previous work at $T_p=0.625$ and
0.8~GeV~\cite{DymovPLB,Kurbatov}, only a hint of reaction (\ref{eq2})
could be seen. For the present energies, a clear peak is revealed
around $M_x^{\,2}\approx 0$ (Fig.~\ref{fig2}). This is well separated
from the $\pi^0$ signal at 0.353~GeV whereas, at 0.5 and 0.55~GeV,
the two structures partially overlap because of broadening of the
pion peak away from the production threshold. Fits of the $M_x^{\,2}$
distributions as the sum of modeled $\gamma$ and $\pi^0$
contributions and a straight line background lead to the parameters
listed in Table~\ref{tab1}. The missing-energy distributions for the
$\gamma$-peak events in the overall c.m.\ frame, which are
reflections of the resolution of the setup and the $E_{pp}$ range,
have widths $\approx1$~MeV. The mean $E_x^m$ in Table~\ref{tab1}
agree with the expected kinematic values to within $\approx 0.2$~MeV.
The energy $E_\gamma$ of the inverse reaction is averaged over the
$E_{pp}$ range $0-3$~MeV and distributed with an rms of 0.5~MeV.

\begin{figure}[t]
\includegraphics[width=0.48\textwidth]{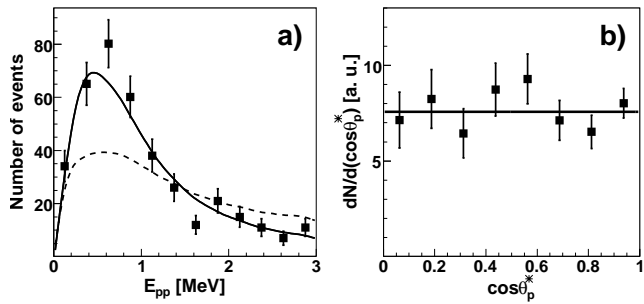}
\caption{Distributions of $pp\to \{pp\}_{\!s}\gamma$ events at
$T_p=0.5$~GeV. a) $E_{pp}$ spectra. The curves represent simulations
with the FSI factor (solid line) and without (dashed). b)
Distribution of acceptance-corrected events for $E_{pp}<3$~MeV with
respect to the cosine of the angle $\theta_p^*$ between the proton
momentum in the diproton rest frame and the diproton momentum in the
overall c.m.} \label{fig3}
\end{figure}

The $E_{pp}$ spectrum of events from the $\gamma$-peak is shown in
Fig.~3a before correcting for acceptance. The solid curve represents
the Monte-Carlo simulation, with events being generated according to
phase space, modified by a Migdal-Watson $pp$ FSI factor taken from
the square of the low-energy $pp$ elastic amplitude in the
$^{1\!}S_0$ wave~\cite{DymovPLB}. These events were traced through
the experimental setup, with due allowance for all its known
features. The simulation satisfactorily reproduces the experiment,
with $\chi^2/\mr{ndf}=11.4/11$ at 0.5~GeV. If the FSI is neglected,
this figure rises to $71/11$. Further evidence that $P$-wave
contamination is at most a few percent is provided by the
acceptance-corrected proton angular distribution in the $pp$ rest
frame. As seen in Fig.~\ref{fig3}b, this is consistent with the
isotropy expected for the $^{1\!}S_0$ final state.

In order to obtain the differential cross section
$d\sigma/d\Omega_{pp}$ as a function of the diproton polar angle
$\theta_{pp}$, events with $E_{pp}<3$~MeV in the $\gamma$-peak of the
$M_x^{\,2}$ distributions were analyzed in $\cos\theta_{pp}$ bins of
$0.01-0.02$ width. After subtraction of the background contamination,
the yield of reaction~(\ref{eq2}) was found from the number of events
in the missing mass intervals $M_x^{\,2} =
0\pm2.5\sigma(M_x^{\,2})_{\gamma}$ at 0.353~GeV and $M_x^{\,2} =
0\pm1.8\sigma(M_x^{\,2})_{\gamma}$ at higher energies. The background
was determined at 0.353~GeV by using missing-mass intervals outside
but adjacent to the $\gamma$ peak. At 0.5 and 0.55~GeV, the
contribution from the tail of the $\pi^{0}$ peak was also considered,
with the shape being taken from the simulation. The event numbers
were corrected for detector efficiency and setup acceptance, as
determined from the full Monte-Carlo simulation.

\begin{figure}[t]
\includegraphics[width=0.48\textwidth]{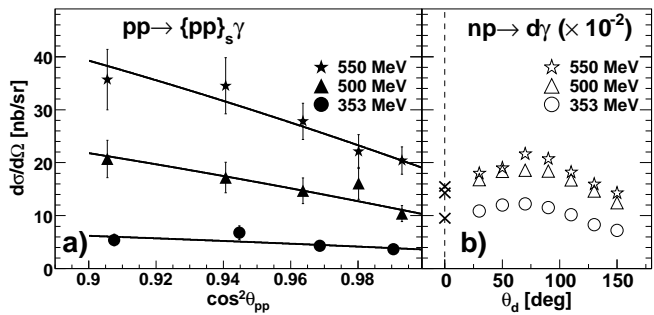}
\caption{a) Angular dependence of the c.m.\ differential cross
section of the $pp\rightarrow\{pp\}_{s}\gamma$ reaction. The errors
shown do not include that of the absolute normalization. The full
curves represent the fits on the basis of Eq.~(\ref{fit}), with the
parameters being given in Table~\ref{tab2}. b) Differential cross
section for the $np\rightarrow d\gamma$ reaction deduced from data on
the inverse reaction~\cite{Crawford}. Crosses at $\theta_d=0^\circ$
are theoretical expectations~\cite{Wilhelm} for energies (down to up)
of 353, 500, 550~MeV.} \label{fig4}
\end{figure}

Since the two initial protons are identical, the differential cross
section is a function only of $x = \cos^2\theta_{pp}$, and the
measured values given in Table~\ref{tab2} indicate a very strong
dependence upon this variable. Theoretical considerations of the
$\gamma\{pp\}_{\!s}\to pp$ reaction~\cite{Wilhelm2} suggest that, in
our energy range, it might be sufficient to retain transitions
corresponding to only the three lowest allowed multipoles, \emph{viz}
$E1$, $E2$ and $M2$. Moreover, it is predicted~\cite{WilhelmPR} that
the $M2$ strength should vanish for $E_\gamma\approx 0.25$~GeV and be
rather low compared to $E1$ and $E2$ in the range $0.18-0.28$~GeV.
Since the $E2$ and $E1$ transitions do not interfere, restricting to
just these two multipoles, the differential cross section is of the
form
\begin{equation}
\label{fit}
\frac{d\sigma}{d\Omega_{pp}}=a[(1+x)\kappa+10x(1-x)]\,,
\end{equation}
where $\kappa = \sigma(E1)/\sigma(E2)$ and $a = 3\sigma(E2)/16\pi$.
Here $\sigma(EJ)$ is the total cross section of reaction (\ref{eq2})
for the $EJ$ multipole. Fitting the data with this form (see
Fig.~\ref{fig4}a) leads to the parameters $\kappa$ and $a$ given in
Table~\ref{tab2}. For all our energies the value of $\kappa$ shows
that the $E1$ and $E2$ multipoles have rather similar strengths, a
feature that was not evident in the $^3{\rm He}(\gamma,pp)n$
experiments~\cite{Emura,Niccolai,Audit}.

The cross sections for the $pp\rightarrow\{pp\}_{\!s}\gamma$ reaction
are compared in Fig.~\ref{fig4}b with those of $np\rightarrow
d\gamma$~\cite{Crawford}. The diproton-to-deuteron ratio is small and
varies with angle and energy between about $4\times 10^{-3}$ and
$3\times 10^{-2}$. In part, this low value is due to the smaller
phase space volume for the unbound $pp$ system and this gives a
suppression factor $\approx 0.1$~\cite{DymovPLB}. The residual
suppression must be related to the different dynamics in the two
reactions. The crucial point here are the absence in the diproton
photodisintegration of the spin-non-flip $E1$ term~\cite{WilhelmPR}
and the $M1$ transition~\cite{Laget}, which dominates the $\gamma
d\rightarrow pn$ reaction in the $\Delta$ energy range. Intermediate
$\Delta N$ states are allowed in $P$ and higher partial
waves~\cite{WilhelmPR}, though their strength will be reduced by the
centrifugal barrier. As a consequence, the contribution of the
$\Delta$ isobar in the $\gamma \{pp\}_{\!s}\rightarrow pp$ absorption
should be greatly diminished compared to the $\gamma d\rightarrow pn$
case. This logic has also been advanced to explain the relatively
small cross section of diproton photodisintegration in the
$^3\mr{He}(\gamma,pp)n$ reaction~\cite{Emura,Niccolai,Audit}.

%
%
\begin{table}[htb]
\caption{Differential cross sections of the $pp\rightarrow
\{pp\}_{\!s}\gamma$ reaction in the c.m.\ system. The errors shown
only take into account statistics and uncertainties in the background
estimation. The overall normalization is known to about $\pm 11$\%.
The fit parameters $\kappa$ and $a$ are defined by Eq.~(\ref{fit}).}
\begin{ruledtabular}
\begin{tabular}{l|cc|cc|cc}
$T_p$ [GeV]&\multicolumn{2}{c|}{0.353} & \multicolumn{2}{c|}{0.500} & \multicolumn{2}{c}{0.550} \\ \hline
&$\theta_{pp}$ & $d\sigma/d\Omega_{pp}$ & $\theta_{pp}$ & $d\sigma/d\Omega_{pp}$ & $\theta_{pp}$ & $d\sigma/d\Omega_{pp}$ \\
&[deg] & [nb/sr] & [deg] & [nb/sr] & [deg] & [nb/sr] \\ \hline
&\phantom{1}5.6  & 3.7$\pm$0.8 & \phantom{1}4.8  & 10.4$\pm$1.5  & \phantom{1}4.8  & 20.4$\pm$2.6 \\
&10.2 & 4.3$\pm$0.9 & \phantom{1}8.1  & 16.1$\pm$3.0  & \phantom{1}8.1  & 22.1$\pm$3.2 \\
&13.6 & 6.8$\pm$1.3 &    11.0 & 14.7$\pm$2.4  &    11.0 & 27.8$\pm$3.4 \\
&17.7 & 5.4$\pm$1.1 &    14.1 & 17.2$\pm$2.9  &    14.1 & 34.5$\pm$5.3 \\
&     &             &    17.9 & 20.7$\pm$3.5  &    17.9 & 35.7$\pm$5.7 \\ \hline
$a$ [nb/sr] & \multicolumn{2}{c|}{$3.1\pm 1.7$}& \multicolumn{2}{c|}{$13.5\pm 4.1$} & \multicolumn{2}{c}{$23.8\pm 6.9$} \\
$\kappa$ &  \multicolumn{2}{c|}{$0.58\pm 0.44$} & \multicolumn{2}{c|}{$0.38\pm 0.16$} & \multicolumn{2}{c}{$0.40\pm 0.16$} \\
\end{tabular}
\end{ruledtabular}
\label{tab2}
\end{table}

%
%
%
%

The most prominent feature in the energy dependence of the $\gamma
d\rightarrow pn$ total and small-angle differential cross sections is
the bump at $E_{\gamma} \approx 150 - 300$~MeV~\cite{Crawford},
caused by the excitation of the $\Delta$-isobar. In contrast, the
total cross section for photon absorption by two bound protons in the
$^3\mr{He}(\gamma,pp)n$ reaction~\cite{Emura} falls steadily as
$E_{\gamma}$ increases from 0.2 to 0.5~GeV, in qualitative agreement
with the arguments for the $\Delta$-isobar suppression. It is also in
line with the results of the model calculation that indicates a
monotonic decrease in the $E2$ contribution through the $\Delta$
region~\cite{WilhelmPR}. Our results are in clear disagreement with
these findings. It is important to note that, if the $M2$ amplitude
is neglected, the parameter $a$ would reflect directly the $E2$
contribution to the $pp\rightarrow\{pp\}_{\!s}\gamma$ total cross
section. The values of $a$ reported in Table~\ref{tab2} rise strongly
with energy and the most plausible explanation for this behavior is
the influence of $D$-wave $\Delta N$ intermediate states.

A rapid rise with angle was also noted in the differential cross
section for single pion production in the $pp\rightarrow
\{pp\}_{\!s}\pi^0$ reaction near the forward
direction~\cite{Kurbatov}. The $S$--wave $\Delta N$ contribution is
also suppressed here by parity and angular momentum conservation,
though a broad maximum was observed in the forward direction at 0.5
-- 1.0~GeV, which might also be a reflection of higher partial waves
in the $\Delta N$ intermediate states.

The $pp\rightarrow\{pp\}_{\!s}\gamma$ analyzing power will also be
measured together with $pp\rightarrow\{pp\}_{\!s}\pi^0$ over an
extended angular range at ANKE by using a polarized proton
beam~\cite{Kulikov}. This is of interest because any signal should
then arise from the interference of the $E2$ with the $E1$ and
$M2$ multipoles~\cite{Wilhelm2}. Even more revealing would be a
measurement of spin correlation with the polarized beam and target
that are available at ANKE~\cite{SPIN}.

An extended study of reaction~(\ref{eq2}), involving also the use of
$\gamma$-detectors, might be feasible at COSY, where the maximum beam
energy is $T_p=2.9$~GeV. An investigation at energies well above the
$\Delta$ region would allow one to compare with other
$^3\mr{He}(\gamma,pp)n$ data~\cite{Niccolai}. The onset of
dimensional scaling, observed at large transverse momenta in $\gamma
d\rightarrow pn$ for $E_{\gamma}> 1$~GeV~\cite{Rossi} and suggested
for $^3\mr{He}(\gamma,pp)n$~\cite{BrodskyPL}, might also be studied
in reaction~(\ref{eq2}).

In summary, the reaction $pp\rightarrow \{pp\}_{\!s}\gamma$ with
production of the final $^{1\!}S_0$ proton pair has been observed at
beam energies of 0.353, 0.50, and 0.55~GeV. The differential cross
sections measured for c.m.\ angles in the interval $0^{\circ} -
20^{\circ}$ are orders of magnitude lower than those for
$np\rightarrow d\gamma$. The rapid change of the $pp\rightarrow
\{pp\}_{\!s}\gamma$ cross section with angle allows one to estimate
the ratio of the $E1$ and $E2$ multipole intensities. The rise in the
differential cross section with energy may be related to the $\Delta$
excitation in higher partial waves. There is no sign of such a
behavior in the data on the photoabsorption on $^3$He, though the
interpretation there is complicated by multinucleon absorption. The
$pp\rightarrow \{pp\}_{\!s}\gamma$ reaction does not suffer from this
drawback and further study of the process, including the measurement
of polarization observables, which is possible at ANKE, might open up
a new way to investigate the properties of the \textit{pp} system at
high momentum transfers.

We are grateful to the COSY machine crew and other members of the
ANKE Collaboration for the support provided. The work was partly
financed by the BMBF Grant to JINR. The external participants
acknowledge the hospitality and support by FZ-J\"ulich. The
interest of Prof.\ S.B.~Gerasimov in the work is also
acknowledged.

\end{document}